\documentclass[epj]{svjour}

\usepackage{graphicx}
\usepackage{psfrag}
\usepackage{fancyhdr}
\def\makeheadbox{{
 }}
\def\mytitle{My title} 
\def\myauthors{My name}  
\def\mytype{My type of session}
\def\mysession{My session}

\def\mytitle{SUSY beyond minimal flavour violation} 
\def\myauthors{Sebastian J\"ager}    
\def\mytype{Review}
\def\mysession{\myauthors}

\newcommand{\eq}[1]{{(\ref{#1})}}

\begin{document}
\title{SUSY beyond minimal flavour violation}
\author{S. J\"ager\inst{1}$^,$\inst{2}
\thanks{\emph{Email:} sebastian.jager@cern.ch}%
\thanks{Talk given at SUSY07, Karlsruhe, Germany, July 26--August 1, 2007}
}                     
%
%
\institute{
Arnold Sommerfeld Center, Department f\"ur Physik,
Ludwig-Maximilians-Universit\"at M\"unchen, Theresienstra{\ss}e 37,
D-80333 M\"unchen, Germany
\and
Theory Division, Physics Department,
CERN, CH-1211 Geneva 23, Switzerland
}
%
\date{}
\abstract{
We review aspects of the phenomenology of the MSSM with non-minimal flavour
violation, including a discussion of important constraints and the
sensitivity to fundamental scales.
} 
\maketitle
%

\section{Introduction}
\label{sec:intro}
The Standard Model (SM) of particle physics provides an economical
description of thousands of observables in particle physics (and
presumably most everyday phenomena) based on the principle of
spontaneously broken gauge invariance. Nevertheless, there are
phenomena which do not find a straightforward explanation within the
SM, notably dark matter, neutrino masses, and the gravitational force.
The latter two, together with the approximate unification of the three running
couplings in the ultraviolet, provide three independent hints at new
dynamics at scales $M_X \sim {\cal O}(10^{15}-10^{19})$
GeV. Explaining and stabilizing the ensuing hierarchy $M_W \ll M_X$
provides some of the principal motivation for supersymmetry, the
eponym of this conference.
SUSY necessarily entails an enlargement of the particle spectrum
(including species suitable to make up the observed dark matter),
naturally improves the unification of couplings, and makes the
hierarchy stable against quantum corrections.

The LHC is designed to directly probe the TeV scale and clarify the
mechanism of electroweak symmetry breaking. If supersymmetry is
involved, ATLAS and CMS will likely detect at least part of the
particle spectrum directly. On the other hand, the examples given above
demonstrate that low-energy observables exist that can probe fundamental
scales, including those beyond the ``energy frontier''. This happens,
for instance, if the fundamental physics violates accidental symmetries of
the low-energy theory, e.g.\ lepton flavour and number in the case of neutrino
oscillations in the context of the seesaw mechanism. SM
contributions to hadronic flavour transitions are constrained by the
chiral structure of weak interactions and suppressed by the weak
scale and the Cabibbo-Kobayashi-Maskawa (CKM) hierarchy,
and, in the case of flavour-changing
neutral current (FCNC) processes, also by loop factors.
Hence these processes, too, provide very sensitive indirect probes.

The remainder of this document is organized as follows. We first review the
Lagrangian of the Minimal Supersymmetric Standard Model (MSSM)
and the anatomy of the effective vertices. Subsequently we consider
some prominent $B$-physics observables where sparticles
may be involved, as well as attempts to combine
various measurements to get detailed information on the Lagrangian.
We close with patterns predicted in specific scenarios of grand
unification.
Throughout, the presentation aims at putting self-containedness over
exhaustiveness.

\section{SUSY flavour violation}
\label{sec:fv}
\subsection{Lagrangian}
\label{sec:lag}
The  MSSM
\cite{Nilles:1983ge,Haber:1984rc,Martin:1997ns}
pairs each SM fermion multiplet with scalar partners into
a chiral superfield and the gauge bosons with gauginos into vector
supermultiplets. The SM Higgs
doublet is replaced by two chiral multiplets $H_u$, $H_d$ (as also
required by supersymmetry), each containing a scalar Higgs doublet.
Imposing $R$-parity to suppress baryon and lepton number violation, the most
general renormalizable superpotential is then fixed by the SM
Yukawa couplings, up to one Higgs mass parameter (the $\mu$-parameter).
In particular, the quartic Higgs couplings are fixed in terms of
the gauge couplings by the supersymmetry, and both supersymmetry and
the electroweak symmetry are unbroken and guaranteed to remain so
at all orders of perturbation theory~\cite{Zumino:1974bg,Grisaru:1979wc}.
The phenomenologically required supersymmetry
breaking can be introduced explicitly as long as it is soft, i.e. does
not introduce quadratic divergences. (In particular, the soft-breaking
terms themselves are only logarithmically sensitive to heavy scales
such as the seesaw scale.) For suitable choices of parameters, the
electroweak symmetry is also broken while the hierarchy $M_W \sim
M_{\rm SUSY} \ll M_X$ is stabilized, where $M_X$ denotes one of the
large scales $M_{\rm seesaw}$, $M_{\rm GUT}$, $M_{\rm Pl}$.
More fundamentally, the relevant extra
renormalizable terms ${\cal L}_{\rm soft}$ in the scalar potential may be
due to spontaneous SUSY breaking, which, if dynamical, can naturally
generate the large hierarchy $M_{\rm SUSY} \ll M_X$ by dimensional
transmutation~\cite{Witten:1981nf}.

Potentially, ${\cal L}_{\rm soft}$ contains a large number of flavour- and
CP-violating parameters. A general parameterization of the
soft-breaking terms is given by a set of trilinear scalar
couplings $T_U$, $T_D$, $T_E$ (counterparts of the trilinear (Yukawa)
couplings in the superpotential), by explicit (generation-dependent) scalar
mass terms for the 5 types of chiral multiplets, explicit gaugino
masses and three parameters in the Higgs sector.
With the exception of the latter two sets, the physical
parameters in ${\cal L}_{\rm soft}$ are fully determined by the $6 \times 6$
hermitian sfermion mass matrices in the so-called super-CKM
basis~\cite{Dugan:1984qf,Hall:1985dx} (of superfields),
\begin{equation}
M^2_{(\tilde u, \tilde d, \tilde e)} =
\left( \matrix{
  M^2_{(\tilde u, \tilde d, \tilde e)  LL} & M^2_{(\tilde u, \tilde d, \tilde e) LR}
\cr \cr
  \big( M^2_{(\tilde u, \tilde d, \tilde e) LR} \big)^\dagger &
	M^2_{(\tilde u, \tilde d, \tilde e) RR} }
\right) ,
\end{equation}
together with the $3 \times 3$ sneutrino mass matrix $M^2_{\tilde \nu}$.
Here,
\begin{eqnarray}
M^2_{\tilde d\, LL} &=& \hat m^2_Q + m_d^2 + D_{d\, LL}, \label{eq:softmat1st}\\
M^2_{\tilde u\, LL} &=& 
        V_{\rm CKM} \hat m^2_Q V^\dagger_{\rm CKM} + m_u^2 + D_{u\, LL}, \\
M^2_{\tilde d\, RR} &=& \hat m^2_{\tilde d} + m_d^2 + D_{d\, RR}, \\
M^2_{\tilde u\, RR} &=& \hat m^2_{\tilde u} + m_u^2 + D_{u\, RR}, \\
M^2_{\tilde d\, LR} &=& v_1 \hat T_D - \mu^* m_d \tan \beta, \\
M^2_{\tilde u\, LR} &=& v_2 \hat T_U - \mu^* m_u \cot \beta,
    \label{eq:softmatlast} \\
D_{f\, LL,RR} &=&
        \cos 2\beta\, M_Z^2 (T^3_f - Q_f \sin^2 \theta_W) {\bf 1}_{3
          \times 3} ,
\end{eqnarray}
with the charged-slepton mass matrices following via the substitutions
$Q \to L$, $d \to e$. Note that the $LR$ masses are
proportional to the electroweak scale, hence are suppressed by
$v/M_{\rm SUSY}$ in the limit of a large SUSY-breaking scale.
The super-CKM basis is defined by requiring diagonal
Yukawa couplings and the CKM matrix being in its four-parameter standard form.
The neutral fermion-sfermion-gaugino and fermion-sfermion-higgsino
couplings are then flavour-diagonal while the charged couplings are governed
by the CKM matrix.
However, the matrices entering \eq{eq:softmat1st}--\eq{eq:softmatlast}
are in general nondiagonal, subject only to certain hermiticity conditions.
The resulting flavour violation is conveniently parameterized in terms
of parameters
\begin{equation}
\label{eq:def_delta}
\big( \delta^{u,d,e,\nu}_{ij} \big)_{AB} \equiv
\frac{ \Big[ \big( {\cal M}_{\tilde u, \tilde d, \tilde e, \tilde \nu}
  \big)_{AB} \Big]_{ij} }{M^2}
\end{equation}
where $M$ is a mass scale of order the sfermion mass eigenvalues.
(A popular flavour-dependent choice is to use
$M^2 = \sqrt{[({\cal M}^2_f)_{XX}]_{ii} [({\cal M}^2)_{YY}]_{jj}}$
in $(\delta^{\tilde f}_{ij})_{AB}$.)
When the $\delta$ parameters are small, it is possible to treat them
as perturbations~\cite{Hall:1985dx}, however caution must be taken
to expand to sufficiently high orders to account for all leading effects (see
also below).

From the large number of flavour-violating parameters, it is evident
that the MSSM generally entails deviations from the SM predictions for
flavour-violating processes. This opens the possibility to either
observe supersymmetry indirectly or to constrain its parameters
(flavour-violating as well as flavour-conserving ones). One virtue of
flavour-violating processes is the large number of observables and the
availability of theoretical tools for rather precise predictions
for many of them. On the other hand, for generic $\delta \sim 1$ and
TeV-scale sparticle masses, experimental bounds on FCNC processed such
as $B \to X_s \gamma$ are violated (the ``SUSY flavour problem'').
However, the structure of ${\cal L}_{\rm soft}$ is
intimately tied to the unknown mechanism of SUSY breaking. For instance,
gauge-mediation models~\cite{Dine:1993yw,Dine:1994vc,Giudice:1998bp}
have little trouble in satisfying the low-energy constraints from
flavour physics because the SUSY breaking is tranferred by
flavour-blind gauge interactions at relatively low scales.

This spectrum of possibilities is exciting: from a phenomenological
point of view, flavour violation provides, as does the sparticle spectrum,
nontrivial constraints on dynamics at fundamental scales.

\subsection{Effective vertices}
\label{sec:vert}
In view of the small masses of $B$, $D$, and light mesons compared to
the weak and SUSY scales, the appropriate tool to separate the heavy
scales from the low-energy QCD effects and curb large logarithms
is the effective weak Hamiltonian (see~\cite{Buras:1998raa} for a
review of the formalism)
\begin{equation}
  \label{eq:heff}
  {\cal H}_{\rm eff} = \sum C_i(\mu) Q_i(\mu),
\end{equation}
where $Q_i$ are local operators constructed from the SM fields,
$C_i$ are Wilson coefficients encapsulating the effects of the sparticles, and
$\mu$ is a renormalization (factorization) scale.
Integrating out the superpartners at the one-loop level, the operators
up to dimension six are determined by penguin and box graphs.

\subsubsection{$\Delta F=2$ (mixing)}
The effective $\Delta F=2$ hamiltonian relevant to meson-antimeson
oscillations is solely due to box diagrams.
A complete operator basis is given by~\cite{Buras:2000if}
\begin{eqnarray}
Q_1^{\rm VLL} &=& (\bar s^a \gamma_\mu P_L b^a) (\bar s^b \gamma^\mu P_L b^b), \\
Q_1^{\rm LR} &=&  (\bar s^a \gamma_\mu P_L b^a) (\bar s^b \gamma^\mu P_R b^b), \\
Q_2^{\rm LR} &=&  (\bar s^a P_L b^a) (\bar s^b P_R b^b), \\
Q_1^{\rm SLL} &=&  (\bar s^a P_L b^a) (\bar s^b P_L b^b), \\
Q_2^{\rm SLL} &=&  - (\bar s^a \sigma_{\mu\nu} P_L b^a)
     (\bar s^b \sigma^{\mu\nu} P_L b^b) ,
\end{eqnarray}
($a,b$ colour indices), plus operators arising from
replacing $P_L \leftrightarrow P_R$ in $Q_1^{\rm VLL}$ and
$Q_{1,2}^{\rm SLL}$. Only $Q_1^{\rm VLL}$ is generated
in the SM (to good approximation).
Supersymmetric contributions have been computed
in~\cite{Gerard:1984bg,Gabbiani:1996hi,Urban:1997gw,Feng:2000kg,Ciuchini:2006dw}.
For illustration, the SUSY gluino-squark contributions in the
mass insertion approximation (expansion in $\delta$'s to lowest
order) read~\cite{Gabbiani:1996hi}
\begin{eqnarray}
   C_1^{\rm VLL} &=& \epsilon   \label{eq:df2mia1st}
      [24 x f_6(x)  + 66 \tilde f_6(x)] (\delta^d_{sb})_{LL}^2 , \\
   C_1^{\rm LR} &=& \epsilon
      [(-12 x f_6(x) - 60 \tilde f_6(x)) (\delta^d_{sb})_{LL} (\delta^d_{sb})_{RR}
\nonumber \\ &&
      + 90 \tilde f_6(x) (\delta^d_{sb})_{LR} (\delta^d_{sb})_{RL}] ,
\\
   C_2^{\rm LR} &=& \epsilon
      [(504 x f_6(x) - 72 \tilde f_6(x)) (\delta^d_{sb})_{LL} (\delta^d_{sb})_{RR}
\nonumber \\ &&
        - 132 \tilde f_6(x) (\delta^d_{sb})_{LR} (\delta^d_{sb})_{RL}], \\
   C_1^{\rm SLL} &=& \epsilon\,
      222 x f_6(x) (\delta^d_{sb})_{RL}^2, \\
   C_2^{\rm SLL} &=& \epsilon
      (-9/2) x f_6(x) (\delta^d_{sb})_{RL}^2, \label{eq:df2mialast}
\end{eqnarray}
where 
$\epsilon = -\alpha_s^2/(216\,m_{\tilde q}^2)$ and three more
coefficients $C_1^{\rm VRR}$, $C_1^{\rm SRR}$, $C_1^{\rm SRR}$
obtained from $C_1^{\rm VLL}$, $C_1^{\rm SLL}$, $C_1^{\rm SLL}$
via $L \leftrightarrow R$.
Here $x = m_{\tilde g}^2/m_{\tilde q}^2$ and $f_6(x)$, $\tilde f_6(x)$
are dimensionless loop functions.

\subsubsection{$\Delta F=1$ (decays)}
QCD-penguin graphs contribute only to $\Delta F=1$
transitions via the operators
\begin{eqnarray}
  Q_{3,4} &=& \sum_q (\bar s_{L} b_{L})\, (\bar q_L q_L) , \\
  Q_{5,6} &=& \sum_q (\bar s_{L} b_{L})\, (\bar q_R q_R) , \\
  Q_{8g} &=& \bar s_{L} b_{R}\, G,
\end{eqnarray}
together with operators $Q_i'$ which arise from the $Q_i$ by changing
the chiralities of all quarks.
Here we have taken the case of $b \to s$ transitions as an example,
with obvious replacements for $b \to d$ and $d \to s$
transitions. We have also suppressed colour and part of the Dirac
structure. ($G$ is the gluon field strength.) Note that the QCD
penguins $Q^{(')}_{3 \dots 6}$ involve $b$ and $s$ quarks of like
chiralities, while the chromomagnetic penguins $Q^{(')}_{8 g}$
involve a chirality flip. This engenders specific patterns of
sensitivity of their coefficients to the parameters $\delta$, e.g.\ 
$(\delta^d_{sb})_{LR}$ in the case of $C_{7\gamma}$ and  $C_{8g}$.

Photon and $Z$ penguins contribute to operators
\begin{eqnarray}
  Q_{9,10} &=&
      \sum_q \frac{3 e_q}{2} (\bar s_{L} b_{L}) (\bar q_L q_L) , \\
  Q_{7,8} &=&
      \sum_q \frac{3 e_q}{2} (\bar s_{L} b_{L}) (\bar q_R q_R) , \\
  Q_{7\gamma} &=& \bar s_{L}\, b_{R} F, \\
  Q_{9V}\, (Q_{10A}) &=&
      (\bar s_{L} b_{L}) (\bar l_L \bar l_L \pm \bar l_R l_R)
\end{eqnarray}
(and their mirror images $Q_i'$; $F$ is the electromagnetic field strength),
which also have analogs contributing to $l_i \to l_j$ transitions.
All the un-primed operators receive significant SM contributions,
which are unimportant for the rest. Tree-level $W$-boson exchange also
generates operators
\begin{equation}
  Q^u_{1,2} = (\bar s_L u_L) (\bar u_L b_L), \quad
  Q^c_{1,2} = (\bar s_L c_L) (\bar c_L b_L) ,
\end{equation}
which do not receive comparable contributions in the MSSM.
Identifying contibutions to some process from any of the primed operators would
constitute a clear signal of new physics.

All penguin operators except the magnetic ones also receive
contributions from box diagrams;
moreover, boxes can contribute to generic four-fermion operators
$$Q = (\bar s_{L,R}\, b_{L,R}) (\bar q_{L,R}\, q_{L,R}) .$$
Note that, in general, boxes are not suppressed with
respect to penguin diagrams~\cite{Gabrielli:1995bd,Gabbiani:1996hi},
as all SUSY penguins as well as boxes decouple
as $M_W^2/M^2$ for $M$ large.
(This is true even for the $Z$-penguin, where the $b s Z$ vertex is
dimension-four but carries a hidden $v^2/M^2$
suppression in its coefficient~\cite{Colangelo:1998pm}.)

Finally, Higgs penguins generating operators such as
$\sum_q y_q (\bar s_{L} b_{R}) (\bar q_L q_R)$
are negligible in the SM. This continues to hold in the MSSM for
small $\tan\beta$. At large $\tan\beta$, depending on the Higgs sector
there is a peculiar phenomenology already for minimal flavour
violation, see e.g.\ 
\cite{Hamzaoui:1998nu,Choudhury:1998ze,Babu:1999hn,Carena:2000uj,Isidori:2001fv,Buras:2002wq,Dedes:2002er,Buras:2002vd,Isidori:2006jh,Isidori:2006pk,Freitas:2007dp,Ellis:2007kb,SusyTrine}.
For detailed studies of large-$\tan\beta$ effects
beyond minimal flavour violation we refer
to~\cite{Foster:2004vp,Foster:2005wb,Foster:2005kb,Foster:2006ze}.
As large $\tan\beta \sim 30-60$ is separately reviewed at this
conference~\cite{SusyIsidori}, in the remainder we will largely
restrict ourselves to small $\tan\beta$.

\section{Phenomenology}
We turn to a discussion of a number of flavour observables that can provide, at
present or in the foreseeable future, potential for constraining or
discovering supersymmetry.
 Schematically, a decay rate provides a constraint
\begin{eqnarray*}
    \lefteqn{|{\cal A}_{\rm SUSY}|^2 + 2\,{\rm Re}\,{\cal A}_{\rm
               SUSY}^* {\cal A}_{\rm SM} }
\\ \qquad         &=& \Gamma_{\rm exp} (1 \pm \Delta^{(\rm exp)})
            - |{\cal A}_{\rm SM}|^2 (1 \pm \Delta^{(\rm SM)}) .
\end{eqnarray*}
The right-hand side often involves a cancellation: the flavour observables
measured so far are consistent with no SUSY contributions.
Hence the goal is precision, reducing the errors $\Delta^{(\rm exp)}$ and
$\Delta^{(\rm SM)}$ as much as possible, while for the
left-hand side the rough dependence on SUSY parameters is enough. Other
observables have similar expressions.
The focus below is on hadronic observables.

\subsection{Mixing}
\subsubsection{$K-\bar K$ mixing}
$K-\bar K$ oscillations played a role in estimating the
charm quark mass before its observation, as well
as in the discovery of (indirect) CP violation, later giving
information on the CP-violating phase in the CKM matrix.
$\Delta M_K$ and $\epsilon_K$ also provide a classic constraint on
supersymmetric flavour (see e.g. \cite{Ellis:1981ts,Barbieri:1981gn}).
The mass difference $\Delta M_K$ and the CP-violating parameter
$\epsilon_K$ follow from the effective $\Delta F=2$ Hamiltonian,
\begin{eqnarray}
  \Delta M_K
&\propto& 
  2 \sum_i B_i\, {\rm   Re}\, C_i , \\
  \epsilon_K &\propto& \frac{e^{i \pi/4}}{\sqrt{2} \Delta M_K}
    \sum_i B_i\, {\rm Im}\, C_i,
\end{eqnarray}
where $B_i \equiv \langle K | Q_i | \bar K \rangle$. The hadronic matrix
elements $B_i$ contain low-energy QCD effects and require
nonperturbative methods such as (numerical) lattice QCD,
see e.g.~\cite{Donini:1999nn,Babich:2006bh,Becirevic:2001xt}.
Moreover, $\Delta M_K$ is afflicted by long-distance contributions
which are believed to be subdominant but difficult to estimate.
Nevertheless, in view of the strong CKM suppression of the SM
contribution, even a rough estimate of the
$B_i$ translates into strong constraints on $s\to d$ flavour violation
parameters, leading to bounds (assuming gluino-squark dominance and
absence of cancellations)
\begin{eqnarray}
|(\delta^d_{ds})_{LL}|, |(\delta^d_{ds})_{RR}| &<& {\cal O}(10^{-2}),
    \label{eq:deltaLLbound}
\\
|(\delta^d_{ds})_{LR}|, |(\delta^d_{ds})_{RL}| &<& {\cal O}(10^{-3}), \\
|(\delta^d_{ds})_{LL} \cdot (\delta^d_{ds})_{RR}| &<& {\cal O}(10^{-7}), 
    \label{eq:deltaLLRRbound}
\end{eqnarray}
for $M \sim m_{\tilde q} \sim m_{\tilde g} \sim 200$ GeV
~\cite{Gabbiani:1996hi,Ciuchini:1998ix,Misiak:1997ei}, a
well-known aspect of the ``SUSY flavour problem''.
The constraints become weaker as the SUSY scale is increased,
scaling roughly like $M$, as is evident from~\eq{eq:df2mia1st}--\eq{eq:df2mialast}.
At any rate, this ``problem'' looks less severe when considering that
the corresponding CKM factor $V_{td}^* V_{ts}= {\cal O}(10^{-4})$ is
also much smaller than its `generic' value ${\cal O}(1)$, and that
the $LR$ $\delta$ parameters are ${\cal O}(v/M)$. Indeed the problem
is completely removed, for instance, in the framework of minimal flavour
violation~\cite{Hall:1990ac,Buras:2000dm,Buras:2000qz,D'Ambrosio:2002ex}.

\subsubsection{$B_d-\bar B_d$ and $B_s-\bar B_s$ mixing}
Here the mixing amplitudes
\begin{equation}
  {\cal A}(\bar B_q \to B_q) \propto M_{12}^q - \frac{i}{2} \Gamma_{12}^q 
\end{equation}
($q=d,s$)
are completely short-distance dominated. Hence the theoretical
expression
\begin{equation}
  \Delta M_{B_q} \propto |M_{12}^q| \sim f_{B_q}^2 | \sum B_i C_i | ,
\end{equation}
where $f_{B_q}$ are decay constants, and $B_i$ again
parameterize hadronic matrix elements,
can be directly compared to the experimental
values~\cite{Barberio:2007cr,Abulencia:2006ze}
\begin{eqnarray}
    \Delta M_{B_q} &=& (0.507 \pm 0.004)\, {\rm ps}^{-1},  \\
    \Delta M_{B_s} &=& (17.77 \pm 0.10 \pm 0.07)\, {\rm ps}^{-1} .
\end{eqnarray}
In both cases, the theory error is fully dominated by $f_{B_q}$. For instance,
$\Delta M_{B_s}^{\rm SM} \approx (16 \dots 27)\, {\rm ps}^{-1}$ seems realistic
depending on the values of $f_{B_s}$~\cite{utfit,ckmfit,tantalo}.
This is consistent with the experimental value (which is
however on the low side). Combined with the fact that the remaining
(perturbative, non-CKM) uncertainties are at the 1--2 percent level,
this underlines the importance of the ongoing efforts to obtain
these nonperturbative parameters on the lattice with a high precision.

On the other hand, the weak phase $\phi_d = \arg M^{d}_{12}$ governs
mixing-decay interference, hence can be extracted cleanly from the
time-dependent CP asymmetry in
$B \to J/\psi K_S$ decay (with theoretical uncertainties
of order $1-2$ \%), giving~\cite{Barberio:2007cr}
$$
    \sin \phi_d = 0.675 \pm 0.026 .
$$
In the SM, $\phi_d = 2 \beta$, but this does not hold in the presence
of new flavour violation.
Constraints analogous to~\eq{eq:deltaLLbound}--\eq{eq:deltaLLRRbound}  
follow from $B_d- \bar B_d$ mixing, see e.g.~\cite{Becirevic:2001jj}.

The mixing phase $\phi_s$ in the $B_s$ system will be measured at LHCb from
the analogous asymmetry in  $B_s \to J/\psi\, \phi$.
In the SM $\phi_s \approx 0$, and any mixing-induced
asymmetry in this mode  would be a crystal clear signal of new physics.

The impact of $B_s-\bar B_s$ mixing data on the
MSSM \cite{Foster:2006ze,Ball:2006xx,Ciuchini:2006dx} is considered
below in subsection~\ref{sec:global}, and in section~\ref{sec:susygut}
in the context of SUSY GUTs.

\subsubsection{$D - \bar D$ mixing}
The observation of $D-\bar D$ oscillations in 2007 at the
$B$-factories~\cite{Aubert:2007wf,Staric:2007dt,Abe:2007rd} provides
a constraint, which is unfortunately
difficult to quantify because the SM contribution to $M_{12}$ is completely
long-distance dominated and rather uncertain. What is
certain is that it has a negligible weak phase, hence mixing-induced CP
violation in $D$ decays would signal non-SM physics. For the time
being, an upper bound on $|M_{12}|$ can be obtained from data and used to
put constraints on up-type $\delta$ parameters analogous to the cases discussed
above~\cite{Ciuchini:2007cw,Nir:2007ac,Golowich:2007ka,Fajfer:2007dy}.

\subsection{Rare $K$ decays}
\begin{figure}
\includegraphics[width=0.45\textwidth,angle=0]{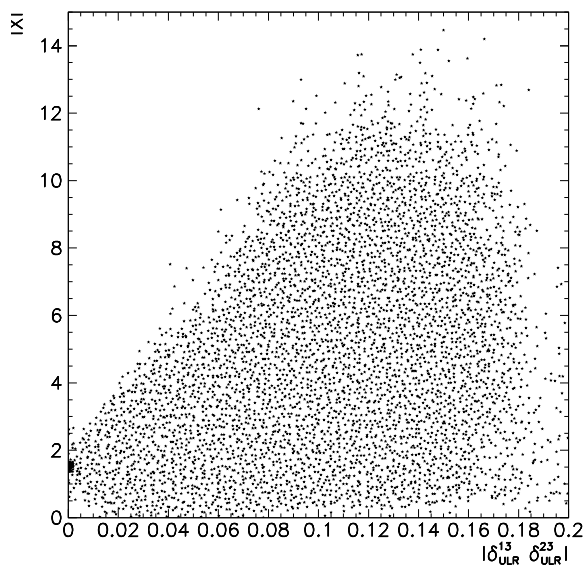}
\caption{SUSY contribution $X$ to ${\cal A}(K \to \pi \nu \bar \nu)$. Figure taken from~\cite{Buras:2004qb}
\label{fig:ampKpinunu}       
}
\end{figure}

The decays $K^+ \to \pi^+ \nu \bar \nu$ and $K_L \to \pi^0 \nu \bar
\nu$ are almost unique in that they are essentially free of hadronic
uncertainties. In the SM context, the two modes provide a clean
and independent determination of the unitarity
triangle~\cite{Buchalla:1994tr,Buchalla:1996fp,Buras:2005gr,Buras:2006gb}---once
they will have been measured precisely, hopefully, at CERN NA48/III and at
JPARC (the SM branching fractions are ${\cal O}(10^{-11})$
and ${\cal O}(10^{-10})$, respectively).
In general, they rather selectivly probe the FCNC vertices of the $Z$
boson. In view of the particular, $SU(2)$-breaking structure of the
leading $Z s d$ vertex, this implies a specific sensitivity to
certain combinations of LR and RL $\delta$-parameters, even in the
presence of general flavour
violation~\cite{Nir:1997tf,Buras:1997ij,Colangelo:1998pm,Buras:1999da,Buras:2004qb,Isidori:2006qy}. In a
perturbative expansion in $\delta$'s,
\begin{equation}
  {\cal A}(K \to \pi \nu \bar \nu)^{\rm SUSY} \propto
    (\delta^u_{dt})_{LR} (\delta^u_{ts})_{RL} ,
\end{equation}
where $(\delta^u_{dt(st)})_{LR}=(\delta^u_{td(ts)})_{RL}^*$ are
related to up-squark (mass)$^2$ matrix elements
in a certain non-super-CKM basis~\cite{Buras:1997ij}. This is an
example where the (generically) leading effect arises at second order
in the mass insertions.
A systematic numerical analysis~\cite{Buras:2004qb}
(see also~\cite{Isidori:2006qy})
shows that this parametric dependence continues to hold beyond the perturbative
expansion, and even in the presence of large contributions from box
diagrams (Fig.~\ref{fig:ampKpinunu}). Moreover the SM hierarchy between the
charged and neutral modes may be reversed, the latter enhanced by an
order of magnitude or more, and the bound following from
isospin~\cite{Grossman:1997sk} saturated.
Complementary probes of the $Z$-penguin amplitude are provided by the
modes $K_L \to  \pi^0 e^+ e^-$ and $K_L \to \pi^0 \mu^+ \mu^-$,
which are still theoretically quite clean~\cite{Isidori:2006qy,Mescia:2006jd}.

\subsection{Leptonic $B$ decays}
Among $B$-decays, the modes $B^+ \to \ell^+ \nu$ and $B_d, B_s \to \ell^+
\ell^-$ are the theoretically cleanest. The former proceeds
through a $W^+$ tree-level diagram in the SM. Significant corrections
may occur due to charged-Higgs-boson exchange, which is present in the
MSSM~\cite{Akeroyd:2003zr,Itoh:2004ye,Isidori:2006pk} but becomes
relevant only at large values of $\tan\beta$. In this case, the latter
mode can be enhanced by an order of magnitude or more, which provides
a serious constraint on certain large-$\tan\beta$ scenarios. (The
branching fraction scales with the sixth power of $\tan\beta$.) At
small $\tan\beta$, it will receive more moderate contributions via
the $Z$-penguin and the operator $Q_{10A}$. In the
SM one has~\cite{Buras:2003td,Haisch:2007ia}
$$
   BR(B_s \to \mu^+ \mu^-) = (3.51 \pm 0.50) \times 10^{-9}, 
$$
where the bulk of the hadronic uncertainties has been eliminated by
normalizing to $\Delta M_s$. (The error will become even smaller with
improved lattice predictions for $\hat B_{B_s}$.) In spite of this
mode being so rare, LHCb, ATLAS, and CMS expect to collect a combined
few hundred SM events after five years or so of running.
Hence this mode will play an important role also at small $\tan\beta$.

\subsection{Semileptonic and radiative decays}
\subsubsection{Inclusive $B \to X_{s} \gamma$}
While being loop-induced, the inclusive decay $B \to X_{s} \gamma$ has
a relatively large branching ratio and is well measured, with
a world average~\cite{Barberio:2007cr} of
$$
   BR(\bar B \to X_s \gamma)_{\rm exp} = (3.55 \pm
   0.24^{+0.09}_{-0.10} \pm 0.03) \times 10^{-4}
$$
(with a 1.6 GeV lower cut on the photon energy).
In the SM, the decay amplitude receives its dominant contributions from
$W-t$ loops entering through the magnetic operator $Q_{7\gamma}$ and $W-c$
loops entering through loop contractions of the tree operators
$Q^c_{1,2}$. Both contributions are of comparable size and opposite in
sign. Other operators are subdominant. The corresponding
state-of-the-art NNLO-QCD prediction reads~\cite{Misiak:2006zs}
(see~\cite{Haisch:2007ic} for a review and discussion of uncertainties)
$$
   BR(\bar B \to X_s \gamma)_{\rm SM} = (3.15 \pm 0.23) \times 10^{-4},
$$
slightly more than $1 \sigma$ below the experiment.
Supersymmetric effects have been investigated
thoroughly in the literature both in minimal
flavour violation~\cite{Bobeth:1999ww,Ciuchini:1998xy,Borzumati:2003rr,Degrassi:2006eh,Degrassi:2000qf,Carena:2000uj,Buras:2002vd,Freitas:2007dp,D'Ambrosio:2002ex}
and beyond~\cite{Borzumati:1999qt,Besmer:2001cj,Ciuchini:2002uv}. 
They enter chiefly through $Q_{7\gamma}$, resulting in a
large sensitivity to the parameter $(\delta^d_{sb})_{LR}$.
(Contributions via $Q'_{7\gamma}$ do not interfere with the SM
contribution in an inclusive process due to the opposite chirality of
the produced $s$-quark, hence generally have a
small impact.) The full one-loop SUSY contribution may
involve a cancellation between charged-Higgs-top loops, which are
always of the same sign as the SM piece, and squark-higgsino as well
as squark-gaugino loops, which (in general) carry an arbitrary complex phase.

\subsubsection{Inclusive $B \to X_{s} \ell^+ \ell^-$}
Their sensitivity to semileptonic operators like $Q_{9 V}$ and $Q_{10 A}$
makes the rare $b \to s \ell^+ \ell^-$ transitions a complementary and
more complex test of the underlying theory than the radiative ones. The
three-body decays allow to study non-trivial observables such as the
dependence on the kinematics  of the decay products. In the absence of
large statistics,
partially integrated spectra such as the dilepton mass spectrum or the
angular distribution can be explored that are amenable to a clean
theoretical description for a dilepton invariant mass below the charm
resonances.

In the minimally flavour-violating  MSSM the Wilson
coefficients $C_{9 V}$ and $C_{10 A}$
are only slightly affected and corrections to the decay
distributions do not exceed the $30 \%$ level~\cite{Ali:2002jg,Bobeth:2004jz}. 
At large $\tan \beta$, additional contributions
to $b \to s \mu^+ \mu^-$ arise from the chirality-flipping operators
$(\bar s_L b_R) (\bar \mu_L \mu_R)$ and $(\bar s_L b_R) (\bar
\mu_R \mu_L)$ that are suppressed by powers of the muon mass but
enhanced by $(\tan \beta)^3$. In practice, these contributions are
however bounded from above~\cite{Chankowski:2003wz,Hiller:2003js,Bobeth:2007dw} by the experimental constraints on
$B_s \to \mu^+ \mu^-$ and turn out to be subleading. Merging the
information on $B \to X_s \ell^+ \ell^-$ with the one on $B \to
X_s \gamma$, one can thus infer that the
sign of the $b \to s \gamma$ amplitude should be
SM-like~\cite{Gambino:2004mv}. In the general MSSM a simultaneous
use of the $B \to X_s \ell^+ \ell^-$ and $B \to X_s \gamma$
constraints leads to stringent limits on
$(\delta^d_{sb})_{LL}$ and $(\delta^d_{sb})_{LR}$ in the complex
plane~\cite{Ciuchini:2002uv,Silvestrini:2007yf}.

\subsection{Combined constraints} \label{sec:global}
\begin{figure}
\includegraphics[width=0.48\textwidth,angle=0]{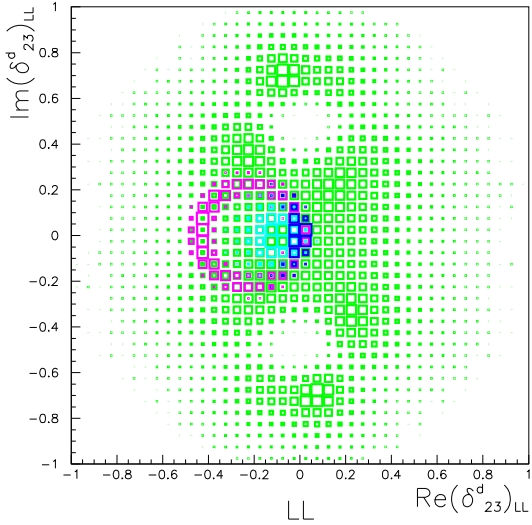}
\caption{Combined constraints on $(\delta^d_{sb})_{LL}$. The allowed
regions due to $B \to X_s \gamma$, $B \to X_s \ell^+ \ell^-$, and
$B_s-\bar B_s$ mixing are shown in violet, light blue, and green,
respectively; the combined constraint is in dark blue. A common
sparticle mass of $350$ GeV and $\mu = -350$ GeV is assumed.
Figure taken from~\cite{Silvestrini:2007yf}
\label{fig:combined} }      
\end{figure}

The constraints on the flavour-violating parameters become more
powerful when the interplay of several observables with different
parametric sensitivities is considered. (This was done in many of
the works referred to above and below.)
For $b \to s$ transitions,
Fig.~\ref{fig:combined} (taken from \cite{Silvestrini:2007yf}) shows 
how the measurements of from $B \to X_s \gamma$, $B \to X_s \ell^+
\ell^-$, and $\Delta M_s$ coact to constrain the parameter 
$(\delta^d_{sb})_{LL}$, leaving a much smaller allowed region than
each individual observable.

\subsection{Charmless hadronic decays}
Two-body exclusive nonleptonic decays $B \to M_1 M_2$ are sensitive to
all of the operators $Q^{(')}_{1\dots10,7\gamma, 8g}$, while offering a
large number ${\cal O}(100)$ of observables, including many CP-violating
ones. They (and exclusive modes in general) also become increasingly
 attractive on the grounds of their accessibility at hadron machines
such as the LHC.
On the theoretical side, the factor limiting the precision are the
hadronic matrix elements $\langle M_1 M_2 | Q_i | B \rangle$, which
involve nonperturbative QCD in a way that is presently not
surmountable in lattice QCD. 
Systematic methods are, however, available, based on expansions
about the limit of $SU(3)$ flavour symmetry or about the
heavy-$b$-quark limit. $m_s/\Lambda$ and $\Lambda/m_b$, respectively,
are the expansion parameters. In fact, the (QCD)
factorization formulae
\cite{Beneke:1999br,Beneke:2000ry,Beneke:2001ev}
for the hadronic matrix elements that follow
from the heavy-quark limit respect the $SU(3)$ flavour symmetry up
to well-defined (so-called ``factorizable'') corrections at the
leading power and perturbative order, and perturbative QCD corrections
do not alter this picture very much. Higher orders in $\Lambda/m_b$
are generally not under control, with certain (important) exceptions.

\begin{figure}
\begin{center}
\includegraphics[width=0.44\textwidth,angle=0]{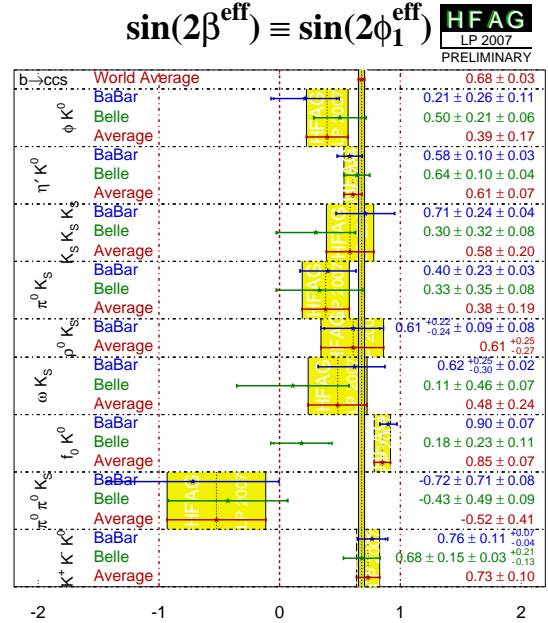}
\caption{Mixing-induced CP violation in charmless $b\to s$
  transitions. Figure taken from~\cite{Barberio:2007cr}}
\label{fig:sin2beff}       
\end{center}
\end{figure}
\subsubsection{Time-dependent CP asymmetries}
One class of observables that has received much interest in recent years
are time-dependent CP asymmetries in decays into CP
eigenstates,
\begin{eqnarray}
\lefteqn{  \frac{BR(\bar B^0(t)\to f) - BR(B^0(t) \to f)}
       {BR(\bar B^0(t)\to f) + BR(B^0(t) \to f)} } \nonumber \\
  & & \equiv S_f \sin(\Delta m_B t) - C_f \cos(\Delta m_B t) .
\end{eqnarray}
In the case of charmless $b \to s$ transitions, within the SM one expects
$ \eta_f S_f \approx \phi_d = 2 \beta $, based on the dominance of the
QCD penguin amplitude, which has vanishing weak phase. ($\eta_f$ denotes the CP
quantum number of the final state.) Neither equality holds in the
MSSM (beyond minimal flavour violation). Fig.~\ref{fig:sin2beff} shows
the data for a number of modes. It is conspicuous that in general, the
$\eta_f S_f \equiv \sin(2\beta^{\rm eff})(f)$ lies below the value of
$\sin 2\beta$.
In fact, in the SM a QCD factorization calculation~\cite{Beneke:2005pu}
of the corrections
for the two-body final states $\phi K^0$, $\eta' K^0$, $\pi^0 K_S$,
$\rho^0 K_S$, $\omega K_S$ due to neglected smaller amplitudes
 shows a (small) positive shift in all cases except $\rho^0 K_S$.
Supersymmetric contributions to $b \to s$ penguin transitions
are well studied, see for instance~\cite{Grossman:1996ke,Lunghi:2001af,Gabrielli:2002fr,Khalil:2002fm,Khalil:2003bi,Agashe:2003rj,Ball:2003se,Kane:2002sp,Kane:2003zi,Harnik:2002vs,Dariescu:2003tx,Ciuchini:2002uv,Chakraverty:2003uv,Khalil:2004yb,Gabrielli:2004yi}.
Often the most important operator is $Q_{8g}$, or even $Q'_{8g}$, which
will interfere with the SM contributions in exclusive processes.
An interesting possibility is to attribute the pattern of the
deviations to constructive and destructive interference between
operators of different quark chiralities, depending on the parities of
the final-state particles~\cite{Khalil:2003bi}. We stress that while
the present data does not appear to show any significant deviations from SM
expectations, the situation in Fig.~\ref{fig:sin2beff} clearly
illustrates the discovery potential of these hadronic observables.

\subsection{Lepton flavour violation}
We do not have time (space) to cover lepton flavour violation in detail, which
is fortunately covered in many other talks at this conference. In the SM,
even when introducing the minimal dimension-five operator to allow for
neutrino masses and mixings, lepton-flavour violating processes such
as $\tau \to \mu \gamma$ are rendered extremely rare by the
tiny neutrino mass splittings and the corresponding near-perfect
GIM cancellation. The situation is very different in the MSSM because
of the presence of lepton flavour violation at the renormalizable
level, in the sneutrino and charged slepton mass matrices. Theoretically, the
SUSY effects in $\ell_i \to \ell_j \gamma$ can be captured in operators
analogous to $Q_{7 \gamma}$ figuring in the discussion of $B \to X_s
\gamma$ above. The experimental upper bounds can likewise be
converted into knowledge about the slepton mass matrices.

\section{Probing the GUT scale} \label{sec:susygut}
\begin{figure}
\begin{center}
\parbox{74mm}{\vskip3mm
\psfragscanon
\psfrag{m6}{$10^{-6}$}
\psfrag{mexp}{$6 \cdot 10^{-7}$}
\psfrag{m6p5}{$10^{-6.5}$}
\psfrag{m7}{$10^{-7}$}
\psfrag{m7p5}{$10^{-7.5}$}
\psfrag{m8}{$10^{-8}$}
\psfrag{m9}{$10^{-9}$}
\psfrag{mgl}{$m_{\tilde g_3}$}
\psfrag{msq}{\Large $m_{\tilde q}$}
\psfrag{adbymsq}[Bl][Bl][1][0]{\Large $a_{d}/m_{\tilde q}$}
\resizebox{74mm}{!}{\includegraphics{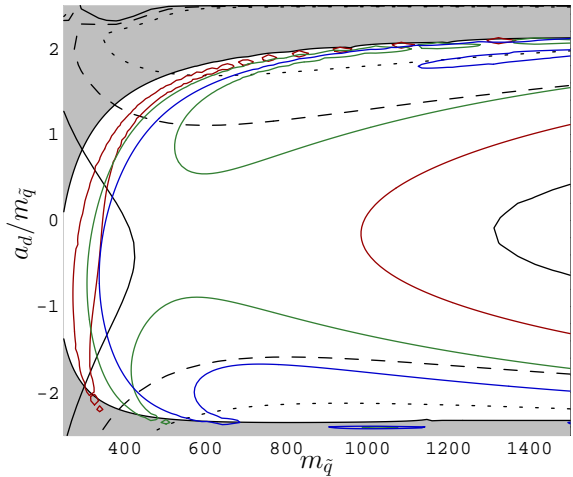}}
}
\end{center}
\caption{Contours of constant SUSY contribution to $\Delta M_{B_s}$,
  in units of the SM value, and of constant
  $BR(\tau\to\mu\gamma)$, in a slice of parameter space.
  Here $a_d$ has a simple relation to $a_0$, $m_{\tilde g} = 250$ GeV,
  $\tan\beta=3$. The solid black, dashed, and dotted contours
  correspond to $|\Delta M_{B_s}^{\rm (NP)}|/\Delta M_{B_s}^{\rm (SM)} = 0.5$,
  $2$, $5$, respectively. The red, green, and blue contours correspond
  to the experimental upper bound on $BR(\tau \to \mu \gamma)$ for
  $\mu=-300$, $\mu=-450$, $\mu=-600$ GeV, respectively. ($\Delta
  M_{B_s}$) is independent of $\mu$ in the approximation used.
  Figure taken from~\cite{Jager:2005ii}}
\label{fig:dmbrtmg}       
\end{figure}

Concrete assumptions about the SUSY-breaking mechanism (gravity
mediation, gauge mediation, etc.) and possible UV completion (such as a SUSY
grand-unified theory, minimal flavour violation, etc.) may imply
patterns in ${\cal L}_{\rm soft}$ relating different $\delta$ parameters
that can be tested against the general constraints applying to them,
or can be further used in making specific predictions for low-energy
observables and their correlations. Recent trends in SUSY models
are reviewed in other talks at this conference~\cite{SusyRaby,SusyNilles}.
One of the most intriguing aspects of SUSY grand unified theories
(GUTs), which also demonstrates the power of flavour
observables to probe even superhigh scales, is the possibility of
relations between hadronic and leptonic flavour violation
(see e.g.\ \cite{Barbieri:1995tw,Barbieri:1995rs,Hisano:1998fj,Baek:2000sj,Moroi:2000mr,Moroi:2000tk,Akama:2001em,Chang:2002mq,Masiero:2002jn,Hisano:2003bd,Goto:2003iu,Ciuchini:2003rg,Jager:2003xv,Jager:2004hi,Jager:2005ii,Grinstein:2006cg,Ciuchini:2007ha,Albrecht:2007ii}).
The effect we are considering here is the
following~\cite{Hall:1985dx,Barbieri:1995tw,Barbieri:1995rs}. Assume
that SUSY breaking is effected at a scale beyond the GUT scale, for
instance at the Planck scale, and that it is
flavour-blind, at least approximately, such that one has a
universal scalar mass parameter $m_0^2$ and a universal $A$-parameter
$a_0$. For definiteness, assume simple $SO(10)$ unification such
that there is only one sfermion multiplet for each
generation. Radiative corrections due to the unified gauge coupling
will correct the masses of the three $16$'s of sfermions in the same
way, while the large top Yukawa coupling will selectively suppress the masses of
one multiplet,
\begin{equation}   \label{eq:so10shift}
    m^2_{16_{1,2}} = m_0^2 + \epsilon, \qquad
    m^2_{16_3} = m_0^2 + \epsilon - \Delta ,
\end{equation}
where $\epsilon \propto g^2$ and $\Delta \propto y_t^2 m_0^2,\, y_t^2 a_0^2$. 
Eq.~\eq{eq:so10shift} holds in a basis where the up-type Yukawa matrix
is diagonal.
(Further contributions will be present if there are additional
large Yukawa couplings, such as for large $\tan\beta$. The expressions
then become more complicated.) In this fashion, the large top Yukawa
coupling affects also the right-handed down-type squark masses and the
slepton masses, which is very different from the situation in the MSSM
(or below the GUT scale). Now the relevant sfermion basis for low-energy
physics (the super-CKM basis) is the one where the down-type and
leptonic Yukawas are diagonal. In such a basis the members of the
$16_3$ selected by $y_t$ will consist of mixtures of the superpartners
of fermions of different flavours, and FCNC SUSY vertices appear.
Moreover, these vertices will be correlated between the hadronic and
leptonic sectors.
This kind of mechanism gained particular attraction after the
observation of large leptonic mixing in neutrino
oscillations~\cite{Chang:2002mq}. Assuming
a minimal-$SU(5)$-type embedding of the MSSM into $SO(10)$, and under
certain assumptions about the generation of seesaw neutrino masses,
one has
\begin{equation}
   Y_D = U_{\rm PMNS}^T\, {\rm diag}(y_d, y_s, y_b)\, V_{\rm CKM}^\dagger.
\end{equation} 
This expression should be most robust in the $(2,3)$ sub-block. The
atmospheric neutrino mixing angle then appears in the gluino-squark
couplings
\begin{equation}
  {\cal L}_{\rm soft} \supset U_{\rm PMNS}^{ij} \, {\tilde d}^*_{Rj} \,
  \tilde g T^A \, d_{Ri} ,
\end{equation}
yielding potentially spectacular effects in observables like
$\Delta M_{B_s}$
\cite{Harnik:2002vs,Jager:2003xv,Jager:2004hi,Jager:2005ii}.
The model is parameterized by four parameters $m_0$, $a_0$, $m_{\tilde
  g}$, and $\mu$, and a value of
$\tan\beta$ around $2-3$ to maintain perturbativity. Fig.~\ref{fig:dmbrtmg}
compares $BR(\tau \to \mu \gamma)$ with the correction to $\Delta
M_{B_s}$. It is evident that the measurement of $\Delta M_{B_s}$
provides a nontrivial and quantifiable constraint on the
lepton-flavour-violating mode, providing (one of many) illustration(s)
of the possibility to probe very fundamental scales with
flavour-violating observables, beyond what is possible with knowing
the particle spectrum alone.

\section{Conclusion}
We have reviewed some of the possible indirect signals of SUSY in
flavour physics and the most important constraints that current
data imposes. We have also discussed some modes and signals that
will be measured in the near future and are likely to be affected by
TeV-scale supersymmetry. Neither set is complete, for instance we did
not have room to discuss exclusive semileptonic and radiative
$B$-decays, which are under active theoretical and experimental investigation.
To close on a positive note, we can hope that soon the era of putting bounds
and constraints will give way to a phase of actual measurements of MSSM
parameters. The interplay between direct and indirect observables
should be useful, as it was in the construction of the SM.
For the longer term, it is then conceivable that
after such a phase indirect probes will take the center stage again.
While it may be the past experience that ``there is interesting physics
at all scales'', the peculiar gentleness of SUSY quantum corrections
may mean that this cannot be extrapolated further. In SUSY, the GUT or Planck
scales can actually be ``close'' as far as indirect observables are
considered (as argued in the previous section), even if being at great
distance from the point of view of direct detection.

\section*{Acknowledgement}
The author would like to thank the organizers for putting together
an exciting conference in a very pleasant environment. I am grateful
to M. Gorbahn and U. Haisch for helpful comments on the manuscript
and to U. Haisch for reminding me of the discussion of unusual signatures of
non-standard penguins in Ref.~\cite{WalGrom}.

%
%

\end{document}